\documentclass[preprintnumbers,amsmath,amssymb,aps,prl,nofootinbib]{revtex4-2}
\newcommand*{\eps}{{\rlap{\lower2ex\hbox{$\,\,\tilde{}$}}{\epsilon_{ijk}}}}
\newcommand*{\EPS}{{\rlap{\lower2ex\hbox{$\,\,\tilde{}$}}{\epsilon_{i'j'k'}}}}
\newcommand*{\lmq}{{\rlap{\lower2ex\hbox{$\,\,\tilde{}$}}{\epsilon_{lmq}}}}
\newcommand*{\jmq}{{\rlap{\lower2ex\hbox{$\,\,\tilde{}$}}{\epsilon_{jmq}}}}
\newcommand*{\jql}{{\rlap{\lower2ex\hbox{$\,\,\tilde{}$}}{\epsilon_{jql}}}}
\newcommand*{\jlm}{{\rlap{\lower2ex\hbox{$\,\,\tilde{}$}}{\epsilon_{jlm}}}}
\newcommand*{\imq}{{\rlap{\lower2ex\hbox{$\,\,\tilde{}$}}{\epsilon_{imq}}}}
\newcommand*{\iql}{{\rlap{\lower2ex\hbox{$\,\,\tilde{}$}}{\epsilon_{iql}}}}
\newcommand*{\ilm}{{\rlap{\lower2ex\hbox{$\,\,\tilde{}$}}{\epsilon_{ilm}}}}
\newcommand*{\lmn}{{\rlap{\lower2ex\hbox{$\,\,\tilde{}$}}{\epsilon_{lmn}}}}
\newcommand*{\abc}{{\rlap{\lower2ex\hbox{$\,\,\tilde{}$}}{\epsilon_{abc}}}}
\newcommand*{\N}{{\rlap{\lower2ex\hbox{$\,\,\tilde{}$}}{N}}}
\newcommand{\tN}{{\rlap{\lower2ex\hbox{$\,\,\tilde{}$}}{N}}}
\newcommand*{\tM}{{\rlap{\lower2ex\hbox{$\,\,\tilde{}$}}{M}}}
\newcommand*{\imn}{{\rlap{\lower2ex\hbox{$\,\,\tilde{}$}}{\epsilon_{imn}}}}
\newcommand*{\qt}{\ln q^{\frac{1}{3}}}

\begin{document}
\title{Intrinsic time gravity,  heat kernel regularization, and emergence of Einstein's theory}

\author{Eyo Eyo Ita III}\email{ita@usna.edu}
\address{Physics Department, US Naval Academy, Annapolis, Maryland}
\author{Chopin Soo}\email{cpsoo@mail.ncku.edu.tw}
\address{Department of Physics, National Cheng Kung University, Taiwan}
\author{Hoi Lai Yu}\email{hlyu@gate.sinica.edu.tw}
\address{Institute of Physics, Academia Sinica, Taiwan}
\input amssym.def
\input amssym.tex

\bigskip\bigskip

\begin{abstract}
The Hamiltonian of Intrinsic Time Gravity is elucidated. The theory describes Schr\"{o}dinger evolution of our universe with respect to the fractional change of the total spatial volume.
Gravitational interactions are introduced by extending Klauder's momentric variable with similarity transformations, and explicit spatial diffeomorphism invariance is enforced via similarity transformation with exponentials of spatial integrals.
 In analogy with Yang-Mills theory, a Cotton-York term is obtained from the Chern-Simons functional of the affine connection. The essential difference is the fundamental variable for geometrodynamics is the metric rather than a gauge connection;
 in the case of Yang-Mills, there is also no analog of the integral of the spatial Ricci scalar curvature. Heat kernel regularization is employed to isolate the divergences of coincidence limits; apart from an additional Cotton-York term, a prescription in which
 Einstein's Ricci scalar potential emerges naturally from the positive-definite self-adjoint Hamiltonian of the theory is demonstrated.
\\
\\
\noindent
Published: E. E. Ita III, C. Soo and H. L. Yu, 2021 Class. Quantum Grav. {\bf 38} 035007
\\
\end{abstract}

\maketitle

\section{Introduction and Fundamentals of intrinsic time gravity}
\label{intro}

Quantum geometrodynamics with first-order Schr\"{o}dinger evolution in intrinsic time, $i\hbar\frac{\delta\Psi}{\delta{T}}={H}_{\rm Phys.}\Psi$, has been advocated in a series of articles\cite{SOOYU,SOOYU1,SOOITAYU,CS,GRwave,second,ORDER}. In this work, we elucidate the Hamiltonian of the theory which contains a Cotton-York term; in addition, heat kernel regularization will be employed to demonstrate the remarkable emergence of Einstein's Ricci scalar potential.\par

In geometrodynamics, the fundamental degrees of freedom are the spatial metric and its conjugate momentum, $(q_{ij},\widetilde{\pi}^{ij})$. The symplectic potential,
\begin{eqnarray}
\label{SYMPLECTIC}
\int {\tilde\pi}^{ij}\delta q_{ij}d^3x = \int ({\bar\pi}^{ij}\delta {\bar q}_{ij} + \frac{1}{3}{\tilde\pi}\delta\ln q)  d^3x,
\end{eqnarray}
\noindent
decomposes in a manner which singles out the canonical pair $(\ln q^{\frac{1}{3}}$, $\widetilde{\pi})$ which commutes with the remaining unimodular metric and traceless momentum variable $(\bar{q}_{ij},\overline{\pi}^{ij})$;
wherein ${\tilde\pi}:=q_{ij}{\tilde\pi}^{ij}$ is  the trace of the momentum, $q$ is the determinant of the spatial metric, and
 \begin{eqnarray}
\label{COMMUTES}
\overline{q}_{ij}:=q^{-\frac{1}{3}}q_{ij};~~\overline{\pi}^{ij}:=q^{\frac{1}{3}}\bigl(\widetilde{\pi}^{ij}-\frac{1}{3}q^{ij}\widetilde{\pi}\bigr).
\end{eqnarray}
\noindent
Hodge decomposition for compact manifolds for the spatial scalar $\delta \qt =\frac{1}{3}\frac{\delta q}{q}$ yields
\begin{eqnarray}
\label{HODGE}
\delta \qt = \delta{T}+\nabla_i\delta{Y}^i,
\end{eqnarray}
\noindent
wherein the spatially-independent $\delta T$ is a 3-dimensional diffeomorphism invariant (3dDI) quantity which serves as the intrinsic time\cite{DeWitt} interval of the theory, whereas $\nabla_i\delta{Y}^i$  is of the form of a  Lie derivative  which can be gauged away by spatial diffeomorphism symmetry.  In fact
\begin{eqnarray}
\label{VOLUME}
{\mathcal L}_{\delta{{\overrightarrow N}}}\qt =\frac{2}{3}\nabla_i{\delta N^i}, \quad \delta T= \frac{2}{3}\frac{\delta V}{V} =\frac{2}{3}\delta\ln V,
\end{eqnarray}
\noindent
wherein $V$ is the spatial volume of the universe which is assumed to be spatially compact\cite{Silk}. In an ever-expanding spatially closed universe, $\delta T$ serves as the preeminent and concrete physical time interval to discuss dynamics and evolution.

A crucial difference between Intrinsic Time Gravity (ITG) and other projectable Horava gravity theories\cite{Horava} (which lack a local Hamiltonian constraint and thus are plagued by a possible extra degree of freedom (d.o.f.)) is that in the latter time is an ``ambient" parameter, whereas in ITG change in a field variable in geometrodynamics, namely $\delta T$, the gauge-invariant change of $\delta \ln q^{\frac{1 }{3}}$, is used as the internal clock interval to deparametrize the theory. The conjugate momentum to $\ln q^{\frac{1 }{3}}$ will be then be proportional to (the negative of) the Hamiltonian density; and the super-Hamiltonian constraint,
\begin{eqnarray}
\label{CONSTRAINT}
\beta^2\tilde{\pi}^2(x)-\bar{H}^2(x)=\beta^2\tilde{\pi}^2(x) -{\bar G}_{ijkl}{\bar \pi}^{ij}(x){\bar \pi}^{kl}(x)-{\cal V}(q_{ij})=0, \qquad {\bar G}_{ijkl} =\frac{1}{2}(\bar{q}_{ik}\bar{q}_{jl}+\bar{q}_{il}\bar{q}_{kj}),
\end{eqnarray}
\noindent
(wherein $\beta$ is a numerical constant\footnote{Variation of $\beta^2:=\frac{1}{3(3\lambda -1)}$ leads to deformation of the DeWitt supermetric $G_{ijkl}:=\frac{1}{2}(q_{ik}q_{jl}+q_{il}q_{kl} -\frac{\lambda}{(3\lambda -1)}q_{ij}q_{kl})$ which has signature $(sgn([\frac{1}{3}-\lambda] ,+,+,+,+,+)$. Einstein's theory corresponds to $\lambda =1$ i.e. $\beta^2=\frac{1}{6 }$, but other values of $\beta$ are allowed in theories which eschew 4-covariance in favor of 3dDI.}) is enforced by eliminating the trace of the momentum in terms of the rest of the
variables. Thus ITG makes use of the super-Hamiltonian constraint to eliminate the one local d.o.f. contained in the pair ($\ln q^{1/3}$, $\tilde\pi$) to yield a reduced physical Hamiltonian\footnote{
The Hamiltonian constraint being quadratic in the trace of the momentum, $\tilde{\pi}$, gives rise to two branches $\beta{\tilde\pi}=\pm{\bar H}$ with $\beta >0$.  Both branches can in fact be investigated within the framework. The classical relation to the trace of the extrinsic curvature is $\tilde\pi = -\frac{\sqrt{q}}{\kappa}K$, so ${\tilde\pi} < 0$  corresponds to an expanding universe, while the other branch corresponds to a universe in contraction.  As will be discussed in the next section, this results in a positive semidefinite quantum Hamiltonian $H_{Phys}=\frac{1}{\beta}\int {\bar H} d^3x$ which is also compatible with an expanding universe. While it is not a necessary condition for stability in exact quantum mechanics, a Hamiltonian bounded from below with stable vacuum is a desirable feature in perturbative quantum field theory.}
\begin{eqnarray}
{H}_{\rm Phys}=-\int {\tilde\pi}(x)d^3x =\frac{1}{\beta}\int {\bar H}(x)d^3x,
\end{eqnarray}
\noindent
which generates evolution in global intrinsic time interval $\delta T$\footnote{A comparison of the reduced Hamiltonian of scalar field time, York extrinsic time and intrinsic time can be found in Ref.\cite{second}}.
 A full canonical analysis of the resultant two physical d.o.f. and the reduced phase space, together with a discussion of the system of second class Dirac constraints, and the associated Faddeev-Popov determinants and functional integral measures has been carried out\cite{second}.
Spatial diffeomorphism invariance is maintained provided the Hamiltonian density,
\begin{eqnarray}
\label{INTRINSIC1}
{\bar H}(x) = {\sqrt{ {\bar \pi^{ij}}  {\bar G_{ijkl}} {\bar \pi^{kl}} + \mathcal{V}(q_{ij})}},
\end{eqnarray}
\noindent
is a scalar density of weight one\footnote{The square-root form of the Hamiltonian density is needed to reproduce the correct equations of motion\cite{SOOYU1,second}. Refs.\cite{ROOT} are literature on the existence and uniqueness of the square root operators, both bounded and unbounded.}.
Einstein's General Relativity (with $\beta= \frac{1}{\sqrt{6}}$ and ${\mathcal V} =- \frac{q}{(2\kappa)^2}[R - 2\Lambda_{\rm{eff}} $]) is thus a particular realization of this wider class of theories.
Horava’s bold proposal\cite{Horava}  to eschew 4-covariance in favor of just spatial diffeomorphism symmetry and its advantages is adopted in ITG in a consistent modification and extension of Einstein’s theory within this framework which yields a hitherto unexplored vista to overcome the many technical and conceptual obstacles in quantum gravity\cite{DeWitt,Wheeler}. It has been shown elsewhere that this Hamiltonian produces the equations of motion of Einstein's theory, with an emergent lapse function which agrees with the a posteriori lapse function of General Relativity\cite{SOOYU,SOOYU1}.

The physical Hamiltonian of the theory, ${H}_{\rm Phys}$, generating translation in cosmic intrinsic time, yields the ordered time-development operator as
\begin{eqnarray}
\label{TIMEDEVELOPMENT}
U(T,T_0)&=&{\bf T}\{{\exp}[-\frac{i}{\hbar}\int^T_{T_0}H_{\rm Phys}(T')\delta{T}']\}\nonumber\\
&=&I-\frac{i}{\hbar}\int^T_{T_0}dT_1H_{Phys}(T_1)
+\Bigl(\frac{-i}{\hbar}\Bigr)^2\int^T_{T_0}dT_2\int^{T_2}_{T_0}dT_1 H_{Phys}(T_2)H_{Phys}(T_1)+\dots\nonumber\\
&&+\Bigl(\frac{-i}{\hbar}\Bigr)^n\int^T_{T_0}dT_n\int^{T_n}_{T_0}dT_{n-1}\dots\int^{T_2}_{T_0}dT_1H_{Phys}(T_n)H_{Phys}(T_{n-1})\dots{H}_{Phys}(T_1) +\dots  .
\end{eqnarray}
In an ever expanding universe, this time-ordering is well-defined and spatially diffeomorphism invariant.

As extensions of the super-Hamiltonian constraint, matter and Yang-Mills interactions can be added under the square root to ${\cal V}$ as $\frac{1}{2\kappa} (H_{matter} +H_{YM})$\cite{second}; for instance, in the case of scalar field, we have the usual
\begin{eqnarray}
\label{SCALAR}
H_\phi(x) = \frac{1}{2}({\tilde\pi}_\phi{\tilde\pi}_\phi + qq^{ij}\nabla_i\phi\nabla_j\phi)+ qV(\phi),
\end{eqnarray}
\noindent
written as a scalar density of weight two, and $H_{YM}$ will be brought up in the next section. The main objective of this current work which shall be elucidated below is to demonstrate the emergence of the Einstein Ricci potential from a Hamiltonian density of the
form ${\bar H}=\sqrt{ {Q}^{\dagger i}_{j}{Q}^{j}_{i}+ q{\cal K}}$.

\section{Quantization and Hamiltonian density}

It is desirable to express the square-root form of the Hamiltonian density as the square-root of a positive-definite and self-adjoint entity.
The traceless momentum variable, $\bar{\pi}^{ij}$, cannot be easily implemented upon quantization as a self-adjoint traceless operator in the metric representation.
Klauder's ``momentric" variable\cite{Klauder}, ${\bar{\pi}}^{i}_{j}$, with one contravariant and one covariant index,  is an intriguing variable which, in lieu of utilizing the momenta, confers many advantages\cite{SOOITAYU}.
Classically,  ${\bar{\pi}}^{i}_{j} =q_{jk}{\bar{\pi}}^{ik}$, as a composite of the spatial metric and momentum the momentric is aptly named. An upshot is the kinetic term in (7) is expressible as  ${\bar \pi^{ij}}  {\bar G_{ijkl}} {\bar \pi^{kl}}={\bar\pi}^i_j{\bar\pi}^j_i$.
The fundamental commutation relations (CR) for the spatial unimodular metric and momentric are,
\begin{eqnarray}
\label{FUNDAMENTALRELATIONS}
&\bigl[\bar{q}_{ij}(x),\bar{q}_{kl}(y)\bigr]=0,\nonumber\\
&\bigl[\bar{q}_{ij}(x),\bar{\pi}^k_l(y)\bigr]=
i\hbar\Bigl(\frac{1}{2}\bigl(\delta^k_i\overline{q}_{lj}+\delta^k_j\overline{q}_{li}\bigr)-\frac{1}{3}\delta^k_l\overline{q}_{ij}\Bigr)\delta(x,y) =: i\hbar\bar{E}^k_{l(ij)}\delta(x,y),\nonumber\\
&\bigl[\bar{\pi}^i_j(x),\bar{\pi}^k_l(y)\bigr]=\frac{i\hbar}{2}\bigl(\delta^k_j\bar{\pi}^i_l-\delta^i_l\bar{\pi}^k_j\bigr)\delta(x,y).
\end{eqnarray}

    Among other advantages, the quantum momentric operator and CR can be explicitly realized in the metric representation
on wavefunctionals of the unimodular metric $\psi[\bar{q}]=\langle\bar{q}_{ij}\vert\psi\rangle$ by functional differentiation
\begin{equation}
{\bar{ \pi}}^{i}_{j}(x)\psi[\bar{q}]=\frac{\hbar}{i}\bar{E}^i_{j(mn)}(x)\frac{\delta\psi[\bar{q}]}{\delta \bar q_{mn}(x)} =\biggl[\frac{\hbar}{i}\frac{\delta}{\delta \bar q_{mn}(x)}\bar{E}^i_{j(mn)}(x)\biggr]\psi[\bar{q}]={\bar{ \pi}}^{\dagger i}_{j}(x)\psi[\bar{q}],
\end{equation}\noindent
with  ${\bar{ \pi}}^{i}_{j}$ being Hermitian on account of $[\frac{\delta}{\delta\bar{q}_{mn}(x)},\bar{E}^i_{j(mn)}(x)]=0$  as well as traceless ($\bar{\pi}^i_i=0$).  Self-adjointness of $\bar{\pi}^i_j$ follows from two properties demonstrated above; namely (i) that $\bar{\pi}^i_j$ is Hermitian; and secondly, given that $\psi[\bar{q}]$ is in the domain of $\bar{\pi}^i_j$,
(ii) that $\psi[\bar{q}]\in Dom(\bar{\pi}^i_j)$ implies that $\psi[\bar{q}]\in Dom({\bar{ \pi}}^{\dagger i}_{j})$ also.

The second CR in (\ref{FUNDAMENTALRELATIONS}) imply that ${\bar\pi}^i_j$ generates infinitesimal $SL(3,R)$ transformations which exponentiates (with traceless $\alpha^i_j$  parameter)  to
\begin{equation}
U^\dagger(\alpha) {\bar q}_{kl}({x}) U(\alpha) = (e^{\frac{\alpha({x})}{2}})^m_k {\bar q}_{mn}({x}) (e^{\frac{\alpha({x})}{2}})^n_l, \qquad e^{\frac{\alpha({x})}{2}} \in SL(3,R), \qquad
 U(\alpha)=e^{-\frac{i}{\hbar}\int \alpha^i_j {\bar \pi}^j_i d^3{y}}.
 \end{equation}
 Thus the momentric generates transformations which preserve the positivity of the eigenvalues of the metric, whereas the momentum variable generates arbitrary translations of the spatial metric which will not in general uphold this positivity\cite{Klauder}.

    Remarkably, the final CR say that momentric operators generate an $su(3)$ algebra at each spatial point
This can be seen from the isomorphism with  $T^A $ through the Gell-Mann matrices  $\lambda^{A=1,...,8}$.  To wit,
\begin{eqnarray}
\label{SU3}
&T^A(x):= \frac{1}{\hbar}(\lambda^A)^j_i{\bar\pi}^i_j (x) \Rightarrow [T^A(x),T^B(y)] =-f^{ABC}T^C\delta(x,y);
\end{eqnarray}
with $f^{ABC}$ being the structure constants of $su(3)$.
The kinetic operator in $\bar{H}^2(x)$ is
\begin{eqnarray}
\label{LAPLACIAN}
\bar{\pi}^j_i(x)\bar{\pi}^i_j(x) \propto T^A(x)T^A(x),
\end{eqnarray}
\noindent
which is self-adjoint and can thus be interpreted as a Casimir invariant of $su(3)$.  This corresponds in
 (\ref{INTRINSIC1}) to $\bar{H}^2(x)$ for the free theory ${\cal V}=0$ with no interactions.\par

 The interactions which are to be introduced must respect the positivity and self-adjointness of $\bar{H}^2(x)$ as well as 3dDI.  To this end, positivity of the Hamiltonian and explicit 3dDI can be achieved by defining the non-Hermitian
operator $Q^i_j$ and its adjoint ${Q^{\dagger}}^i_j$ via
\begin{eqnarray}
\label{WEHAVE}
{Q}^{i}_{j}&= e^{W_T}{{\bar \pi}}^{i}_{j}e^{-W_T}=\frac{\hbar}{i}{\bar E}^i_{j(mn)}[\frac{\delta}{\delta{\bar q}_{mn}} - \frac{\delta W_T}{\delta{\bar q}_{mn}}];~~
{Q^{\dagger}}^{i}_{j}&= e^{-W_T}{{\bar \pi}}^{i}_{j}e^{W_T}=\frac{\hbar}{i}{\bar E}^i_{j(mn)}[\frac{\delta}{\delta{\bar q}_{mn}} + \frac{\delta W_T}{\delta{\bar q}_{mn}}],
\end{eqnarray}
\noindent
and constructing the Hamiltonian density as
\begin{eqnarray}
\label{SELFADJOINT}
\nonumber\\
\bar{H}(x) = \sqrt{ {Q}^{\dagger i}_{j}(x){Q}^{j}_{i}(x)+ q(x){\cal K}},
\end{eqnarray}
\noindent
with positive coupling ${\cal K}$ for the bare cosmological constant allowed by the symmetries of the theory.  Thus, interactions are introduced by extending the momentric with similarity transformations, and explicit spatial diffeomorphism invariance is enforced via similarity transformation with exponentials of spatial integrals, wherein
\begin{eqnarray}
\label{WTOTAL}
W_T=\alpha W_{EH} + g W_{CS}=\alpha\int {\sqrt q}Rd^3x+\frac{g}{4}\int{\tilde\epsilon}^{ijk}({\Gamma}^l_{im} \partial_j{\Gamma}^m_{kl} +\frac{2}{3}{\Gamma}^l_{im}{\Gamma}^m_{jn}{\Gamma}^n_{kl})\,d^3x.
\end{eqnarray}
\noindent
Here $W_{EH}$  is the spatial Einstein-Hilbert action with coupling constant of inverse length dimension, and $W_{CS}$ the Chern-Simons functional of the affine connection, $\Gamma^i_{jk}$, of the spatial metric\cite{YACKIW}, with dimensionless coupling
constant $g$.  It is possible to consider theories with higher curvature invariants in $W_T$; a brief analysis of the effects will be presented in the remarks section.

The main calculation which follows will focus on $W_T$ of Eq.(\ref{WTOTAL}). This yields
\begin{eqnarray}
\label{Qoperator}
&{Q}^{i}_{j}
=\frac{\hbar}{i}{\bar E}^i_{j(mn)}\frac{\delta}{\delta{\bar q}_{mn}} +i\alpha\hbar\sqrt{q}{\bar R}^i_j   + ig\hbar{\tilde C}^i_j,\nonumber\\
&{Q^{\dagger}}^i_j=\frac{\hbar}{i}{\bar E}^i_{j(mn)}\frac{\delta}{\delta{\bar q}_{mn}} -i\alpha\hbar\sqrt{q}{\bar R}^i_j   - ig\hbar{\tilde C}^i_j;
\end{eqnarray}
wherein ${\bar R}^i_j= R^i_j- \frac{1}{3}\delta^i_j R$ is the traceless part of the spatial Ricci tensor\footnote{Due to the ${\bar E}^i_{j(mn)}$ projector only the traceless part of the Ricci tensor ${\bar R}^i_j$  survives in (18); and
any volume term $\lambda\int {\sqrt q}d^3x$ in $W_T$ commutes with ${{\bar \pi}}^{i}_{j}$.}; and
  the Cotton-York tensor\cite{YORK} density is expressible, remarkably, as ${\tilde C}^i_j =\bar{E}^i_{j(mn)}\frac{\delta W_{CS}}{\delta \bar q_{mn}}$. That it is traceless (${\tilde C}^i_i =0$), and transverse ($\nabla_i {\tilde C}^i_j=0$) can be established from
\begin{eqnarray}
\label{COTTON}
\tilde{C}^{ij}=\frac{\delta W_{CS}}{\delta q_{ij}} = \epsilon^{imn}\nabla_m (R^j_n -\frac{1}{4}R\delta^j_n) =
\frac{1}{2}(\epsilon^{imn}\nabla_m R^{j}_n +\epsilon^{jmn}\nabla_mR^{i}_n);
\end{eqnarray}
and the last equality holds because the antisymmetric part of the intermediate entity vanishes on account of the Bianchi identity, $\nabla_m (R^m_n -\frac{1}{2}R\delta^m_n) =0$.  In three dimensions,  the manifold is conformally flat iff the Cotton--York tensor vanishes identically\cite{YORK}.

 It is also of relevance to note that the appearance of the Cotton-York term is entirely natural. It has the analogy of the magnetic field term in Yang-Mills Hamiltonian which can in similar fashion be introduced by extending the momentum conjugate to the gauge connection, ${\hat{\tilde \pi}^{ia}_A}$  by similarity transformation with the exponential of the Chern-Simons functional, $W_{CS}[A]$, of the Yang-Mills connection $A_a =A_{ia}dx^i$, so that
\begin{eqnarray}
\label{YANGMILLS1}
W_{CS}[A] := \frac{1}{2}\int (A^adA_a +\frac{1}{3}f_{abc}A^a\wedge A^b\wedge A^c), \qquad {\hat Q}^{ia} = e^{W_{CS}[A]}{\hat{\tilde \pi}^{ia}_A}e^{-W_{CS}[A]} = {\hat{\tilde\pi}^{ia}_A} +i\hbar{\tilde B}^{ia},
\end{eqnarray}
\noindent
and similarly for its adjoint ${\hat{Q^{ia}}^{\dagger}}$.  The second equality above follows from the conjugate momentum to the gauge potential and the magnetic field being, respectively,
\begin{eqnarray}
\label{YANGMILLS2}
{\hat{\tilde\pi}^{ia}_A} =\frac{\hbar}{i}\frac{\delta}{\delta A_{ia}},\qquad
{\tilde B}^{ia} = \frac{\delta W_{CS}[A]}{\delta A_{ia}}.
\end{eqnarray}
 In Yang-Mills theory, the Hamiltonian density is thus expressible as
\begin{eqnarray}
\label{YANGMILLS}
H_{YM} = \frac{1}{2}q_{ij}( {\hat{\tilde\pi}^{ia}_A}{\hat{\tilde\pi}^{ja}_A}+ \hbar^2{\tilde B}^{ia} {\tilde B}^{ja} )=\frac{1}{2}q_{ij}{{\hat Q}^\dagger}\,^{ia}{\hat Q}^{ja},\qquad q_{ij}[{\hat{\tilde \pi}^{ia}_A},{\tilde B}^{ja}]=0,
\end{eqnarray}
\noindent
with the Yang--Mills squared electric field playing the role of the kinetic and and the squared magnetic field playing the role of the potential term.
\noindent
As explained, the Cotton-York tensor density in (\ref{COTTON}) is the functional derivative w.r.t. the spatial metric of the Chern-Simons functional $W_{CS}$ of the affine connection. The essential difference with Yang-Mills theory is that the fundamental variable for geometrodynamics is the metric rather than a gauge connection.
 In addition, in the case of Yang-Mills theory, there is no analog of the spatial Einstein-Hilbert action term, $\int {\sqrt q}Rd^3x$,  which can be added to $W_T$.  The Einstein--Hilbert term $W_{EH}$, is the term which will eventually lead, through regularization, to the Einstein Ricci scalar potential in the Hamiltonian density.\par
\indent
To summarize, instead of postulating `detailed balance' terms as in Ref.\cite{Horava},  the interactions can be thought of as being introduced through $Q^i_j$ (via similarity transformation of the momentric) with exponential of the Chern-Simons functional associated with dimensionless coupling constant (as in Yang-Mills theory) and the additional spatial Einstein-Hilbert action available to geometrodynamics. It should also be remarked that introduction of interactions via similarity transformations, or conjugation, has been carried out  before\cite{Witten}.

It follows from the above discussion and (\ref{Qoperator}) that  the Hamiltonian density is
\begin{eqnarray}
\label{com}
\bar H &=& \sqrt{ {Q}^{\dagger i}_{j}{Q}^{j}_{i}+ q{\cal K}}\cr
&=& \sqrt{{\bar{\pi}}^{\dagger j}_i {\bar{\pi}}^{i}_j+\hbar^2(g\tilde{C}^i_j + \alpha\sqrt{q}{\bar R}^i_j)(g\tilde{C}^j_i +\alpha\sqrt{q}{\bar R}^j_i ) +i\alpha\hbar \sqrt{q}[{\bar{\pi}}^{i}_j,{\bar R}^j_i] +q{\cal K}}\,.
\end{eqnarray}
Moreover, it can be verified explicitly that $[{\bar{\pi}}^i_j,\tilde{C}^j_i]=0$  (analogously $q_{ij}[{\hat{\tilde \pi}^{ia}_A},{\tilde B}^{ja}]=0$) , so there is no Cotton-York contribution in the commutator term in (\ref{com}).

While ${Q}^{\dagger i}_{j}$ and ${Q}^{i}_{j}$ are related to ${\bar{\pi}}^i_j$ by $e^{\mp W_T}$  similarity transformations, they are non-Hermitian and generate two unitarily inequivalent representations of the non-compact group $SL(3,R)$ at each spatial point; whereas the momentric  ${\bar{\pi}}^i_j = \frac{1}{2}({Q}^{\dagger i}_{j} +{Q}^{i}_{j})$  is self-adjoint and thus generates a unitary representation of ${\prod}_{x }SU(3)_x$. In fact, the commutator term in  ${\bar H}$ can be identified as
\begin{equation}
 i\alpha\hbar{\sqrt q}[{\bar{\pi}}^{i}_j,{\bar R}^j_i] =\frac{1}{2}[ {{Q}^\dagger}^{i}_{j}, {Q}^{j}_{i}].
\end{equation}
 In the functional Schrodinger representation, the commutator at coincident spatial position can be expressed as
\begin{equation}
\label{ZERO2}
\frac{1}{2}[{{Q}^\dagger}^{i}_{j}(x), {Q}^{j}_{i}(x)]= \alpha\hbar^2\hbox{lim}_{x\rightarrow{y}}\sqrt{q(x)}\bar{E}^i_{j(kl)}(x)\frac{\delta\bar{R}^j_i(y)}{\delta\bar{q}_{kl}(x)}.
\end{equation}
\noindent
Using the following formula for the infinitesimal variation of the Ricci tensor and the following useful relation involving the projector
\begin{eqnarray}
\label{RICCIVARIATION}
\delta R_{ac} = \frac{1}{2}q^{bd}(\nabla_b\nabla_a\delta q_{cd}+ \nabla_b\nabla_c\delta q_{ad} -\nabla_b\nabla_d\delta q_{ac}- \nabla_a\nabla_c\delta q_{bd});\nonumber\\
\frac{\partial{\bar q}_{ab}}{\partial {\bar q}_{cd}}= \frac{1}{2} (\delta^c_a\delta^d_b + \delta^c_b\delta^d_a) -\frac{1}{3}{\bar q}_{ab}{\bar q}^{cd}
\end{eqnarray}
\noindent
wherein ${\bar q}^{cd}$ is the inverse unimodular metric, the point-split result is
\begin{equation}
\label{ONE}
\bar{E}^i_{j(kl)}(x)\frac{\delta\bar{R}^j_i(y)}{\delta\bar{q}_{kl}(x)} = -(\frac{5}{6}\nabla^2 + \frac{5}{3}R)\delta(x-y).
\end{equation}
Thus
\begin{equation}
\label{ZERO9}
\frac{1}{2}[{{Q}^\dagger}^{i}_{j}(x), {Q}^{j}_{i}(y)] =-c_1\alpha\hbar^2\sqrt{q(x)}(c_2\nabla^2_x+R)\delta(x-y);  \qquad c_1 = \frac{5}{3}, c_2 =\frac{1}{2}.
\end{equation}
The $x =y$ coincident limit is divergent, and requires regularization which will be addressed  below.

\section{Heat kernel regularization, and emergence of Einstein's theory}
The heat kernel, $K(\epsilon;x,y)$ with $ \hbox{lim}_{\epsilon\rightarrow{0}}K(\epsilon;x,y)=\delta(x-y)$, presents  the means to regularize
the coincident limit in (\ref{ZERO9}) for generic metrics. It satisfies the heat equation,
\begin{equation}\nabla^2K(\epsilon;x,y)=\frac{\partial K(\epsilon;x,y)}{\partial\epsilon};\end{equation}
and in terms of Seeley-DeWitt coefficients $a_n$, $2\sigma(x,y)$ the square of the geodesic length, and $\Delta_V$ the Van Vleck determinant,
\begin{eqnarray}
\label{ZERO10}
K(\epsilon;x,y)=(4\pi\epsilon)^{-\frac{3}{2}}\Delta_V^{\frac{1}{2}}(x,y)e^{-\frac{\sigma(x,y)}{2\epsilon}}\sqrt{q}\sum_{n=0}^{\infty}a_n(x,y;\nabla^2)\epsilon^n,
\end{eqnarray}
wherein $\epsilon$ is of dimension $L^2$.   In the coincidence $x=y$  limit, the coefficients for closed manifolds are\cite{DV}
\begin{eqnarray}
\label{ZERO11}
a_0=1;~~a_1=\frac{R}{6};~~a_2=\frac{1}{180}(R^{ijkl}R_{ijkl}-R^{ij}R_{ij})+\frac{1}{30}\nabla^2R+\frac{R^2}{72};  ...
\end{eqnarray}

The coincidence limit of $ \hbox{lim}_{\epsilon\rightarrow{0}}K(\epsilon;x,y)=\delta(x-y)$ is a regularization of $\delta(0)$, with

\begin{eqnarray}
\label{HEATKERNELL}
K(\epsilon; x,x) =\sqrt{q}(4\pi)^{-3/2}\Bigl(a_0\epsilon^{-3/2} +a_1\epsilon^{-1/2}+a_2\epsilon^{1/2}+ \dots\Bigr)
\end{eqnarray}
\noindent
and as regularization of $\nabla^2\delta(0)$ we may take the coincidence limit of the heat equation,
\begin{eqnarray}
\label{HEATKERNELL1}
\nabla^2\delta(0)
=\hbox{lim}_{\epsilon\rightarrow{0}}\left[\nabla^2 K(\epsilon;x,x)=\frac{\partial{K}(\epsilon;x,x)}{\partial\epsilon}\right]
\end{eqnarray}
\noindent
To wit, at the level before removal of the regulator we have
\begin{equation}
\label{ZERO16}
\frac{\partial{K}(\epsilon;x,x)}{\partial\epsilon}
=\sqrt{q}(4\pi)^{-\frac{3}{2}}\Bigl(-\frac{3}{2}\epsilon^{-\frac{5}{2}}-\frac{a_1}{2}\epsilon^{-\frac{3}{2}}+\frac{a_2}{2}\epsilon^{-\frac{1}{2}}+\dots\Bigr)
\end{equation}
 as a Laurent series. Substituting, as regularizations, $ \hbox{lim}_{\epsilon\rightarrow{0}}K(\epsilon;x,x)$  for $\delta(0)$ and $\hbox{lim}_{\epsilon\rightarrow{0}}\frac{\partial{K}(\epsilon;x,x)}{\partial\epsilon}$ for $\nabla^2\delta(0)$, the regularized coincident limit of (\ref{ZERO9}) takes the form
\begin{eqnarray}
\label{ZERO17}
&\frac{1}{2}[{{Q}^\dagger}^{i}_{j}(x), {Q}^{j}_{i}(x)]=\hbox{lim}_{\epsilon\rightarrow{0}}(-\alpha c_1\hbar^2\sqrt{q})\biggl[c_2\frac{\sqrt{q}}{(4\pi)^{\frac{3}{2}}}
\Bigl(-\frac{3}{2}\epsilon^{-\frac{5}{2}}-\frac{a_1}{2}\epsilon^{-\frac{3}{2}}+\frac{a_2}{2}\epsilon^{-\frac{1}{2}}+\dots\Bigr)\cr
&+\frac{\sqrt{q}}{(4\pi)^{\frac{3}{2}}}\Bigl(\epsilon^{-\frac{3}{2}}a_0+a_1\epsilon^{-\frac{1}{2}}+\dots\Bigr)R\biggr]\nonumber\\
&=\hbox{lim}_{\epsilon\rightarrow{0}}\frac{(-\alpha c_1\hbar^2q)}{(4\pi\epsilon)^{\frac{3}{2}}}\Bigl[-\frac{3c_2}{2\epsilon}+\frac{(12-c_2)R}{12}
+ (a_1 R -c_2\frac{a_2}{2})\epsilon +{\rm terms\,with\,\epsilon^{n\geq 2}}\Bigr];
\end{eqnarray}
wherein $\frac{12-c_2}{12}R=  a_0R -\frac{c_2}{2}a_1 $.
This yields the Hamiltonian density
\begin{eqnarray}
\label{ham}
\bar H &=&\sqrt{ {Q}^{\dagger i}_{j}{Q}^{j}_{i}+ q{\cal K}}\\ \nonumber
&=&\textstyle{\hbox{lim}_{\epsilon\rightarrow{0}}}
\Big\{
{\bar{\pi}}^{\dagger j}_i {\bar{\pi}}^{i}_j+\hbar^2(g\tilde{C}^i_j + \alpha\sqrt{q}{\bar R}^i_j)(g\tilde{C}^j_i +\alpha\sqrt{q}{\bar R}^j_i ) \\ \nonumber
&&+q\Big({\cal K} +\frac{3\alpha c_1c_2\hbar^2}{16\pi^{\frac{3}{2}}{\epsilon^{\frac{5}{2}}}}\Big)-\frac{\alpha c_1\hbar^2q}{(4\pi\epsilon)^{\frac{3}{2}}}\Big(\frac{12-c_2}{12}R+ (a_1 R -c_2\frac{a_2}{2})\epsilon\Big)
\Big\}^{\frac{1}{2}}.
\end{eqnarray}

In the theory there are three fundamental coupling constants: $g, {\cal K}$ and $\alpha$. In the limit of regulator removal with ${\epsilon\rightarrow{0}}$, the divergent $\epsilon^{-\frac{3}{2}}$ term can be countered by $\alpha$.
To yield the correct Newtonian limit, the renormalized finite value is identified phenomenologically, with
\begin{eqnarray}
\label{RENORMALIZED}
\frac{1}{(2\kappa)^2} =\frac{\alpha c_1\hbar^2}{(4\pi\epsilon)^{\frac{3}{2}}}(\frac{12-c_2}{12})
\longrightarrow\alpha =\frac{12(4\pi\epsilon)^{\frac{3}{2}}}{c_1\hbar^2(12-c_2)(2\kappa)^2}
\end{eqnarray}
\noindent
and finiteness of $\kappa$ implies ${\alpha\rightarrow{0}}$ as ${\epsilon\rightarrow{0}}$.  Furthermore, it is noteworthy that the theory actually produces all the Seeley-DeWitt coefficients $a_n$, but these higher curvature terms associated with higher powers of $\epsilon$ in (\ref{ham}) all disappear upon regulator removal, leaving behind just the first Einstein Ricci scalar term with the $-\frac{1}{(2\kappa)^2}$  coupling constant.\par
\indent
The naive effective cosmological constant $\Lambda_{\rm eff}$ can be read off from
\begin{eqnarray}
\label{RENORMALIZED1}
\frac{2\Lambda_{\rm eff}}{(2\kappa)^2} = {\cal K} +\frac{3\alpha c_1c_2\hbar^2}{16\pi^{\frac{3}{2}}{\epsilon^{\frac{5}{2}}}}.
\end{eqnarray}
\noindent
 For positive ${\cal K}$, this naive value of $\Lambda_{\rm eff}$ diverges upon regulator removal; however, zero point energies (ZPE) of other bosonic and fermionic fields (which have, respectively, positive and negative ZPE) have not been taken into account, and they all contribute to the net cosmological constant.  In the $SU(3)\times SU(2)\times U(1)$ Standard Model with bosonic Higgs and gauge fields, and {\it three generations of fermions}, it is well-known that there are ``{\it too many fermions}", resulting in the overabundance of net {\it negative} ZPE. With all ZPE, together with the contribution from a positive ${\cal K}$ taken into account, the net renormalized $\Lambda_{\rm eff}$ may well be positive and finite by countering with ${\cal K}$.

 It should be noticed that from a mathematical point of view it is possible to choose $\alpha \propto \epsilon^{\frac{5}{2}}$, so that no divergences would appear in the limit of vanishing $\epsilon$ and with finite ${\cal K}$. Consequently, there would also be no Einstein Ricci scalar and other curvature terms, and apart from the cosmological constant contribution, only the Cotton-York term remains. Following our earlier arguments on the introduction of interactions, it is noteworthy that this would be, for geometrodynamics, the counterpart of renormalizable Yang-Mills theory with dimensionless coupling constant. Such a theory however would be missing the Einstein scalar curvature term and its physical effects (and it would fail to yield the correct equations for gravitational waves, as discussed in Ref.\cite{GRwave}).
  In the earlier and adopted prescription, starting from the proposed Hamiltonian to its regularization, an interesting relation and physical motivation emerges: that keeping the cosmological constant contribution divergent at this stage (by keeping $\alpha/\epsilon^{\frac{3}{2}}$ finite) retains just the Einstein term and also yields the opportunity to counter net {\it negative} Standard Model zero point energy with divergent positive $({\cal K} +\frac{3\alpha c_1c_2\hbar^2}{16\pi^{\frac{3}{2}}{\epsilon^{\frac{5}{2}}}})$.  However, this does not explain why the resultant $\Lambda_{\rm eff}$ from the cancellation would naturally be as small as the current empirical value\footnote{Expansion of fields in modes and creation-annihilation operators in elementary free fermionic quantum field theory leads to the Hamiltonian
  $H_F =\sum_\sigma\int p^0(a^\dagger_{{\bf p},\sigma}a_{{\bf p},\sigma}-b_{{\bf p},\sigma}b^\dagger_{{\bf p},\sigma}) d^3p
  =\sum_\sigma\int p^0(a^\dagger_{{\bf p},\sigma}a_{{\bf p},\sigma}+b^\dagger_{{\bf p},\sigma}b_{{\bf p},\sigma}-\delta^3(0)\delta_{\sigma\sigma'})d^3p$, wherein $\{b_{{\bf p},\sigma}, {b^\dagger}_{{\bf p'},\sigma'}\}_+ = \delta^3({\bf p}-{\bf p’})\delta_{\sigma\sigma'}$ has been invoked (see, for instance, Ref.\cite{Weinberg}).
  A similar expression with bosonic $+a a^\dagger$ replacing the $–b b^\dagger$ term occurs for bosonic field obeying commutation, instead of anti-commutation, relations. Thus $\pm p^0\delta^3(0)$ zero point energies of opposite signs appear when creation operators of bosonic, respectively fermionic, fields are commuted to the left of annihilation operators with the same eigenvalue. In our Hamiltonian density, expansion of the fields in terms of modes and associated creation-annihilation operators are expected to lead to similar  divergences when creation and annihilation operators with the same eigenvalues are commuted. Supersymmetry curbs the problem by having a bosonic counterpart (with zero point energy of opposite sign) to each fermionic field. With three generations and an overabundance of fermions, the Standard Model will have net negative zero point energy. The upshot is that instead of an infinite {\it negative} ${\cal K}$ which is needed to render the resultant $\Lambda_{\rm eff}$ in Eq.(38) positive, incorporation of Standard Model fields can lead to ${\cal K}> 0$ overcoming the excess negative fermionic contribution to yield a net $\Lambda_{\rm eff} > 0$.
  }.

The final Hamiltonian in the limit of regulator removal in (\ref{ham}) becomes
\begin{eqnarray}
\label{FINAL}
&{H}_{\rm Phys}=\frac{1}{\beta}\int {\bar H}(x) d^3x, \\ \nonumber
&\bar H = \sqrt{ {Q}^{\dagger i}_{j}{Q}^{j}_{i}+ q{\cal K}}
=\sqrt{{\bar{\pi}}^{\dagger j}_i {\bar{\pi}}^{i}_j+g^2\hbar^2\tilde{C}^i_j\tilde{C}^j_i - \frac{q}{(2\kappa)^2}(R - 2\Lambda_{\rm eff})}.
\end{eqnarray}
Apart from the extra Cotton-York potential, Einstein's theory emerges wonderfully, and in a whole new light.

\section{Further remarks}

In $W_T$, higher curvature invariants can be considered; for instance the addition of $\gamma \int \sqrt{q} R^2 d^3x $ with $\gamma$ of length dimension to $W_T$ would lead to an extra term proportional to $i\hbar \gamma R{\bar R}^i_j$  on the R.H.S of Eq.(18).  Carrying through the same analysis, the net effect is that the R.H.S. of (35) contains additional terms of the form
\begin{equation}
\gamma q\hbar^2[\frac{c' _0 R}{\epsilon^{5/2}} + \frac{ c'_1(``R^2" terms)}{\epsilon^{3/2}} + \frac{ c'_2 (``R^3"  \rm{terms})}{\epsilon^{1/2}} +{\rm terms\,with\,\epsilon^{n\geq 0}}].
\end{equation}
This leads to phenomenological identification of finite Newton's gravitational constant which implies
\begin{eqnarray}
\label{FINITE}
\frac{1}{(2\kappa)^2} =\frac{\alpha c_1\hbar^2}{(4\pi\epsilon)^{\frac{3}{2}}}(\frac{12-c_2}{12}) +\frac{c' _0\gamma }{\epsilon^{5/2}}\longrightarrow\alpha \sim \frac{\epsilon^{3/2}  }{(2\kappa)^2},~~\gamma \sim \frac{\epsilon^{5/2}  }{(2\kappa)^2}
\end{eqnarray}
\noindent
and so in the limit of regulator removal ${\epsilon\rightarrow{0}}$, the final form of the Hamiltonian as in (39) is unaffected. This procedure and conclusion do not apply (because no term proportional to just the Einstein $R$ will appear) when even higher curvature invariants (e.g. $\int \sqrt{q} ``R^{n\geq 4}"d^3x$ ) are introduced into $W_T$; consequently the final form of the Hamiltonian will include higher curvature terms which will differ phenomenologically from Einstein's theory. For the case of $n=3$, explicit computations reveal that $\int {\sqrt q}R^i_jR^j_kR^k_i d^3x$  and $\int {\sqrt q}RR^i_jR^j_i d^3x$  do produce a term linear in $R$, whereas $\int {\sqrt q}R^3 d^3x$  does not.
As stated in the beginning of the article, within the context of 3dDI, Einstein's General Relativity is a particular realization of a wider class of theories. It is consistent to have generic ${\cal V}(q_{ij})$ in the framework of Dirac's second class constraints\cite{second}; and similarity transformations of the momentric with higher curvature invariants in $W_T$  are generalizations compatible with the semi-positive-definite  requirement of ${\bar H}^2$ in the Hamiltonian. It should be noted that in $W_T$  the Einstein-Hilbert invariant with cosmological constant, $W_{EH}$, is the lowest order curvature term, and $W_{CS}$, which produces the Cotton-York contribution (${\tilde C}^i_j$ has three spatial derivatives) and comes with dimensionless coupling g,  is the next order term which comes {\it before} other higher curvature invariants.  In the attempt to reconcile unitarity with renormalizability, the availability of the Cotton-York tensor was crucial to Horava’s advocacy of forgoing 4-covariance in favor of just spatial diffeomorphism symmetry and higher spatial derivatives in the Hamiltonian\cite{Horava}.

In the limit of large volume and long wavelengths, the Cotton--York term in (39) will be negligible compared to $qR$ which would would dominate; this corresponds to the familiar Einstein-Hilbert era at late times provided $\beta =\sqrt{\frac{1}{6}}$.
At the opposite extreme of vanishing volume or at the earliest intrinsic times with $q\rightarrow{0}$, the $qR$ term will be small compared to the Cotton-York potential since ${\tilde C}^i_j$ is invariant under scaling and independent of $q$.
A natural question is what governs physics in this era, known as the Big--Bang era.  With the availability of the Cotton--York term,  the upshot is an initial Cotton-York dominated era\cite{SOOITAYU} associated with a quantum vacuum state which satisfies $Q^i_j|0\rangle=0$.
Since ${Q}^{i}_{j}= e^{gW_{CS}}{\bar \pi}^{i}_{j}e^{-gW_{CS}}$, it follows that the wave function is of the form $\langle \bar q_{ij}|0\rangle \sim e^{gW_{CS}}$, which has critical points precisely at the vanishing of the Cotton-York tensor (which is the statement of spatial conformal flatness). Classically, it is intriguing that the lowest energy state is one of vanishing ${\bar\pi}^{ij}$ and vanishing ${\tilde C}^{ij}$, which is phase space initial data compatible with Robertson-Walker spacetimes and Penrose's Weyl Curvature Hypothesis\cite{Penrose}.

While Sakharov's induced gravity from matter loops \cite{SAKAROV} can produce similar emergent terms, the origin of the terms in this work is very different indeed. The basis of the work here has a simple analogy. In the harmonic oscillator, the Hamiltonian is $H=  (a^\dagger a + \frac{1}{2})\hbar\omega$, wherein
\begin{equation}a^\dagger a = \frac{m\omega}{2\hbar}q^2 +\frac{1}{2m\omega\hbar}p^2 -\frac{1}{2}; \qquad a \equiv    \sqrt{\frac{m\omega}{2\hbar}}q + i\sqrt{\frac{1}{2m\omega\hbar}}p, \quad a^\dagger \equiv    \sqrt{\frac{m\omega}{2\hbar}}q -i\sqrt{\frac{1}{2m\omega\hbar}}p.
\end{equation}
Without the reassuring  L.H.S., it is not easy to surmise that the R.H.S is actually positive-definite; moreover, the negative term which contributes to the ZPE of $+\frac{1}{2}\hbar\omega$ in the Hamiltonian actually arose from
$ -\frac{1}{2}=\frac{i}{2\hbar}[q,p]=\frac{1}{2}[a^\dagger, a] $.  The analog in (\ref{ZERO9}) has a similar origin, but unlike the  case of simple harmonic oscillator in quantum mechanics, the commutator at coincident point in a quantum field theory needs regularization, with the consequent emergence of Einstein's scalar curvature from such a framework.
\\
\section{Acknowledgements}
This work has been supported in part by the US Naval Academy, the  Ministry of Science and Technology (R.O.C.) under Grant No. MOST 109-2112-M-006-001, and the Inst. of Physics, Academia Sinica.

\begin{thebibliography}{}
%
%

\bibitem{SOOYU} {C. Soo  and H. L. Yu, Prog. Theor.  Exp. Phys. (2014)  013E01.}

\bibitem{SOOYU1} {N. O' Murchada,  C. Soo and H. L. Yu,   Class. Quantum Grav. {\bf 30} (2013) 095016.}

\bibitem{SOOITAYU} {E. Ita, C. Soo and H. L. Yu, Prog. Theor. Exp. Phys. (2015)  083E01.}

\bibitem{CS} {C. Soo, Int. J. Mod. Phys. D{\bf  25}  (2016) 1645008.}

\bibitem{GRwave} {E. Ita, C. Soo and H. L. Yu, Eur. Phys. J. C (2018) {\bf 78}: 723.}

\bibitem{second} {E. Ita, C. Soo and H. L. Yu, Phys. Rev. D{\bf 97} (2018) 104021.}

\bibitem{ORDER} {C. Soo, Int J. Mod. Phys. D{\bf 25} (2016) 16450008}.

\bibitem{DeWitt} B. S. DeWitt, Phys. Rev. {\bf 160} (1967) 1113. In this work, local intrinsic time $\propto q^{\frac{1}{4}}(x)$ was introduced. Multi-fingered time presents path-dependent ordering problem which does not arise in the
global x-independent time parameter adopted in this article.

\bibitem{Silk}For observational evidence of our compact universe, see,  for instance,  E. Di Valentino, A. Melchiorri  and J.  Silk,  Nat. Astron. {\bf 4} (2020) 196.

\bibitem{Horava} P. Horava,  Phys. Rev. D {\bf 79}  (2009)  084008.

 \bibitem{Wheeler}J. A. Wheeler, {\it Superpsace and the Nature of Quantum
Geometrodynamics}, in Battelle Rencontres, 1967 Lectures in
Mathematics and Physics, edited by C. M. DeWitt and J. A. Wheeler
(W. A. Benjamin, New York, 1968).


\bibitem{ROOT}S. J. Bernau, {\it The square root of a positive self-adjoint operator}, J. Austr. Math. Soc., {\bf 8} (1968) 17-36; A. Wouk, {\it A note on square roots of positive operators}, SIAM Rev. {\bf 8} (1966) 100-102; Z. Sebestyen and Z. Tarcsy, {\it Characterizations of self-adjoint operators}, Studia Sci Math. Hunger. {\bf 50} (2013) 423-435.

\bibitem{Klauder}J. R. Klauder, Int. J. Geom. Meth. Mod. Phys. {\bf 03} (2006) 81, and references therein.

\bibitem{YACKIW}See, for instance,  R. Jackiw, {\it Fifty years of Yang-Mills theory and my contribution to it}, arXiV:physics/0403109; S. Deser, R. Jackiw, and S. Templeton, Phys. Rev. Lett.  {\bf 48} (1982) 975.

\bibitem{YORK}E.  Cotton,  (1899). {\it Sur les varietes a trois dimensions}, Annales de la Faculte des Sciences de Toulouse. II. {\bf  1} (4): 385–438; J. W. York, Jr., Phys. Rev. Lett. {\bf 26} (1971) 1656–1658;  R. Arnowitt, S. Deser, C. W. Misner: {\it The  Dynamics  of  General  Relativity}, in {\it Gravitation: An Introduction to Current Research}, L. Witten (Ed.) ( Wiley, New York, 1962); for a review of the properties of the Cotton tensor, see, for instance, A. Garcia, F. W. Hehl, C. Heinicke, A. Macias, Class. Quantum Grav. {\bf 21} (2004) 1099-1118.

\bibitem{Witten} See, for instance, E. Witten, J. Diff. Geom. {\bf 17} (1982) 661, which featured a Hamiltonian of the form $H= d_g d^\dagger_g + d^\dagger_g d_g $  which is constructed by extending the Laplacian  via $d_g= e^{-gh({\vec\phi})} d e^{gh(\vec{\phi})}$ with Morse function $h(\vec\phi)$.

\bibitem{DV} {D. V. Vassilevich,  Phys. Rept. {\bf 388} (2003) 279.}


\bibitem{Penrose}R. Penrose, {\it Singularities and Time-Asymmetry},  in S. W. Hawking,  W. Israel (eds.), in {\it General Relativity: An Einstein Centenary Survey} pp. 581–638  (Cambridge University Press, 1979).

\bibitem{SAKAROV} {A. D. Sakarov, Dokl. Acad. Nauk SSSR 177 (1967) 70-71}.

\bibitem{Weinberg}Steven Weinberg, {\it The Quantum Theory of Fields}, Vol. 1 (Cambridge University Press, Cambridge, UK, 1995).














\end{thebibliography}

\end{document}